\documentclass[11pt]{article}

\usepackage[T1]{fontenc}
\usepackage[utf8]{inputenc}
\usepackage{lmodern}
\usepackage[margin=1.15in]{geometry}
\usepackage{setspace}
\usepackage[round]{natbib}
\usepackage{booktabs}
\usepackage{graphicx}
\graphicspath{{figures/}}
\usepackage{caption}
\usepackage{flafter}
\usepackage[hidelinks]{hyperref}

\title{\bfseries Mapping Diplomatic Representation in Europe, 1648--1715\\[0.45em]
{\large\mdseries New Data from the \emph{Repertorium der diplomatischen
Vertreter}}}
\author{Magnus Lundgren\\ \small University of Gothenburg\\ \small magnus.lundgren@gu.se}
\date{June 2026\\ \small Working paper}

\begin{document}

\null\vfill

\begin{center}
{\LARGE\bfseries Mapping Diplomatic Representation in Europe,\\[2pt]
1648--1715\par}
\vspace{0.7em}
{\large New Data from the \emph{Repertorium der diplomatischen Vertreter}\par}
\vspace{2em}
{\large Magnus Lundgren\par}
\vspace{0.3em}
{\small University of Gothenburg\\
magnus.lundgren@gu.se\par}
\vspace{1.3em}
June 2026\\
{\small Working paper}
\end{center}

\vspace{2.4em}

\begin{abstract}
\noindent
This paper introduces new data on diplomatic representation in Europe between
1648 and 1715, drawn from Band~I of the \emph{Repertorium der diplomatischen
Vertreter aller L\"ander}. The data comprise 13,344 diplomatic missions, exchanged among 141 sending and
201 receiving polities, and 8,852 individual representatives. The paper
describes the source and coding procedure, assesses the quality and limits of the
data, and reports patterns in participation, rank, and mission duration. In an
illustrative application, I use the data to examine diplomatic continuity across
ruler successions, with findings suggesting that individual appointments remained
tied to the person of the ruler even as the relationships they served proved
durable. The dataset will be useful for research on recognition and membership in
the states system, status and hierarchy among polities, and the formation of the
early modern state. Data collection will continue, and Bands~II and III will
extend the series to 1815.
\end{abstract}


\vfill

\newpage

\section{Introduction}

Diplomacy is among the oldest institutions of the international system. Its
modern European form emerged with the resident embassies of Renaissance Italy
and became general after the Peace of Westphalia \citep{mattingly1955,
anderson1993}. Scholars have long studied how this system took shape, including
the diffusion of resident representation and the ordering of states by rank and
precedence \citep[e.g.,][]{osiander2001, watkins2008}. These accounts rest
predominantly on qualitative evidence, including histories of particular courts,
envoys, and negotiations. Quantitative data on diplomatic representation, by
contrast, exist from 1817 \citep[the Correlates of War Diplomatic Exchange and
H-DATA series;][]{bayer2006, teorell2023}, with later datasets adding annual or
sampled observations for the nineteenth and twentieth centuries
\citep{moyer2021, niklassontowns2023}. These data have supported a wide-ranging
scholarship in international relations, including work on status and recognition
\citep{renshon2016, duque2018}, on the formation of diplomatic
networks \citep{kinne2014}, and on representation and trade \citep{rose2007}.
However, for the preceding two centuries, during which enduring practices and
institutions of the international system in Europe emerged and solidified, no such
data exist.

This paper introduces a dataset of diplomatic representation in Europe covering
1648--1815, built from the \emph{Repertorium der diplomatischen Vertreter aller
L\"ander}, a three-volume reference work compiled from European foreign-ministry
archives in the early twentieth century \citep{bittnergross1936}. This version
covers 1648--1715. It records 13,344 individual diplomatic missions exchanged
among 233 distinct political authorities, and names 8,852 representatives. These
authorities include sovereign monarchies, republics, free imperial cities,
ecclesiastical principalities, imperial diets, and powers outside Christian
Europe such as the Ottoman Empire and Persia. Each mission records the sending
polity,\footnote{I use \emph{polity} throughout as a convenient label for any
authority that sent or received diplomatic representation; some, such as the
imperial diets and assemblies, are not polities in a strict sense.} the receiving
polity, the representative and the rank held, and any event of the mission (such
as the presentation of credentials, an audience, or a departure). For most
missions, dates are given to the day, giving the data a very high temporal
granularity. The data are compiled at the mission level, but I also arrange them
into an annual dyad-year panel.

Descriptive analysis of the aggregate data reveals three patterns. First, at
the level of the system, diplomatic activity was already extensive by the
mid-seventeenth century and grew thereafter: around 200 missions were under way
in the late 1640s, rising to between 550 and 700 by the 1660s and remaining at
that level through the major wars between 1672 and 1714. Second, at the level of
the polity, participation was broad but asymmetric. The Habsburg court at Vienna
received about one mission in ten, while dozens of smaller polities sent or
received only a handful. Third, at the level of the mission, standing residencies
made up about a fifth of representation, and this share did not rise over the
period.

To showcase the utility of the data, I examine the impact of ruler successions
on a polity's diplomatic representation network. This allows me to shed light on
a central question in the literature on state formation \citep{nww2009,
ertman1997}, namely whether the diplomatic apparatus of the early modern state
was patrimonial, an extension of the ruler's own person, or bureaucratic and
impersonal. Two results emerge. First, diplomatic missions were sensitive to
ruler successions, suggesting that appointments remained tied to the ruler.
Second, ruler successions were not associated with a change in a polity's
diplomatic partners, suggesting that its international network was shaped by
relationships more durable than the individual appointments.

These data should be useful for several lines of inquiry that have so far been
pursued mainly with qualitative evidence, or in quantitative work only for the
period after 1815. A first concerns recognition and membership of the states
system. Since the dataset records the full range of polities that exchanged
diplomatic representatives, it can be used to study which polities were admitted
to diplomatic society, extending an established area of research back to the
early modern era \citep{kinne2014, teorell2023, osiander2001}. A second concerns
status and hierarchy, for which the data on diplomatic rank provided here offer a
behavioral measure of precedence among states \citep{duque2018, renshon2016,
bruneau2023}. A third concerns the character of the early modern state, and the
debate over whether its institutions remained dynastic and patrimonial, organized
around the person of the prince, or had begun to acquire an impersonal,
bureaucratic form, a question touched on in the illustrative application below
\citep{nexon2009, teschke2003, ertman1997}. A fourth is diplomatic history, where
these data should enable broader and comparative study of questions relating to
diplomats as individuals, to courts as sites of politics, and to the diplomatic
corps as an emerging profession \citep[e.g.,][]{watkins2008, sowerbyhennings2017}.

The remainder of the paper proceeds as follows. Section~2 describes the source,
data-construction procedures, and coverage. Section~3 presents descriptive
patterns at the system, polity, and mission levels. Section~4 examines
diplomatic continuity following ruler succession. Section~5 concludes.

\section{Data}

\subsection{Source: the Repertorium}

The main source of the dataset is the \emph{Repertorium der diplomatischen
Vertreter aller L\"ander seit dem Westf\"alischen Frieden}, a reference work
that records the diplomatic representatives exchanged across Europe after the
Peace of Westphalia. Band~I, edited by Ludwig Bittner and Lothar Gro{\ss},
appeared in 1936 and covers the period 1648--1715 \citep{bittnergross1936}.
Bands~II and III, edited by Friedrich Hausmann \citeyearpar{hausmann1950} and
Otto Friedrich Winter \citeyearpar{winter1965}, continue the record to 1815.
This paper uses Band~I, while Bands~II and III will be added in subsequent
versions.

The data collection appearing in the \emph{Repertorium} was undertaken under
the auspices of the International Committee of Historical Sciences as a
decade-long international collaboration funded by the Deutsche
Forschungsgemeinschaft, the Rockefeller Foundation, and the Austrian government
\citep{harsin1942}. More than sixty correspondents contributed to the first
volume, each compiling a national chapter from the state and foreign-ministry
archives of the country concerned.

Band~I, comprising about 800 pages, is organized into chapters, each covering a
sending polity (e.g., France) and, within each chapter, by receiving polity,
with missions ordered by their starting date. Entries identify the
representative and report rank, dated acts marking the course of the mission, and
an archival or published source; many also record the mission's purpose. The
volume covers permanent and special missions, including ambassadors, envoys,
residents, agents, commissioners, and treaty plenipotentiaries.

\subsection{Units of observation and inclusion rules}

The elementary record is the printed entry, a block of text devoted to one
representative under one sending polity and one receiving polity. An entry
reports one or more dated acts, each a documented event in the course of a
representation, such as the presentation of credentials, an audience, a recall,
or a departure. The principal unit of analysis is the mission, i.e., the
representation of one polity to one receiver by one or more representatives,
delimited by its first and last recorded act. An entry is coded as a mission when
it records at least one qualifying act, meaning an act of diplomatic
representation other than the signing of a treaty alone.

Some entries record only a treaty signature and no other act. These are excluded
from the main data but retained in a separate set of treaty signings. An entry
that records both a treaty signature and another qualifying act appears in both
sets. On this coding, Band~I yields 13,344 missions and 1,498 treaty signings.

\subsection{Extraction and coding}

The volume was digitized as a scanned image with a machine-readable text layer
produced by optical character recognition (OCR).\footnote{I am grateful to the
University Library, University of Gothenburg, for assistance in this process.} A
layout-aware parser, implemented in Python, converts this text into structured
records, one entry at a time, using the typography to divide each entry into
component fields, including the sending polity, the receiving polity, the
representative, the rank, the dated events, and the source. Chapter headings mark
the sending polity, subheadings the receiving polity, and spacing and type size
separate names, entries, and footnotes.

\begin{figure}[!ht]
\centering
\includegraphics[width=.92\textwidth]{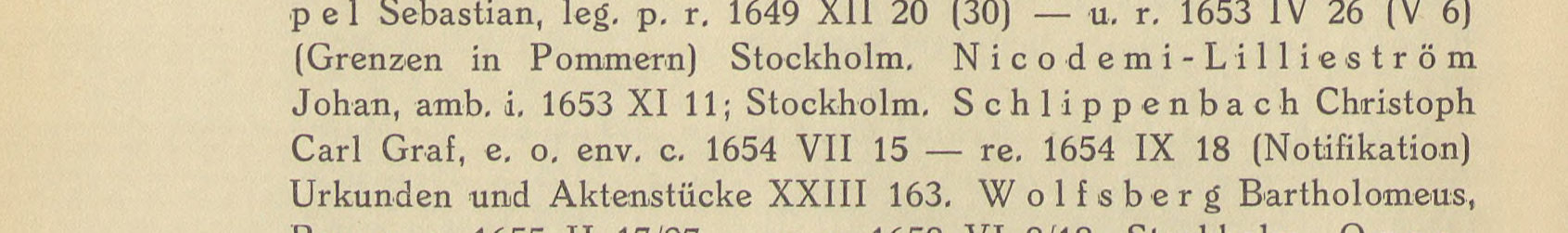}
\caption{A passage of the Swedish chapter as printed (Band I, p.~482). The
entry beginning ``S\,c\,h\,l\,i\,p\,p\,e\,n\,b\,a\,c\,h'' is parsed in
Table~\ref{tab:example}.}
\label{fig:entry}
\end{figure}

Figure~\ref{fig:entry} shows a passage from the chapter on Sweden.
Table~\ref{tab:example} gives the record extracted from its first entry. The
entry names Christoph Carl von Schlippenbach, an envoy sent by Sweden to
Brandenburg-Prussia in 1654, and records the dates on which his credentials were
issued and accepted. The printed entry is terse: the rank appears as
``e.~o.~env.'', short for \emph{envoy\'e extraordinaire}, and the two acts as
``c.'' (credentials, 15 July 1654) and ``re.'' (recredentials, 18 September
1654), with the month in Roman numerals and the purpose (\emph{Notifikation}, a
notification) in parentheses. These abbreviations are the Repertorium's own,
defined in the volume's key to rank and act codes. The parser expands each
occurrence against that key, coding the rank along the five variables shown in
Table~\ref{tab:example} and each act as a controlled event type; where the volume
gives a date in both the Old Style (Julian) and New Style (Gregorian) calendars,
it records both. The same procedure is applied to every entry in the volume.

\begin{table}[!ht]
\centering\small
\caption{The parsed record for the entry in Figure~\ref{fig:entry}.}
\label{tab:example}
\begin{tabular}{ll}
\toprule
Field & Value\\
\midrule
Mission id & 11402\\
Sending polity & Schweden (Sweden)\\
Receiving polity & Brandenburg-Preu{\ss}en (Brandenburg-Prussia)\\
Representative & Schlippenbach, Christoph Carl, Graf\\
Rank, as printed & e.~o.~env.\\
Rank, coded & ceremonial level: envoy or minister;\\
 & extraordinary: yes; resident post: no;\\
 & full powers: no; papal series: no\\
Event & credentials issued, 15 July 1654\\
Event & recredentials accepted (\emph{re.}), 18 September 1654\\
Purpose & Notifikation\\
Source & Urkunden und Aktenst\"ucke XXIII 163\\
Page & 482\\
\bottomrule
\end{tabular}
\end{table}

A mission is treated as active from its first to its last recorded event. The
Repertorium reports events with very high temporal granularity. The opening event
is dated to the day for 91 percent of missions, to the month for 4 percent, and
to the year for the remaining 5 percent. Both the opening and closing events are
dated to the day for 61 percent of missions.

\subsection{Variables and dataset structure}

The core of the dataset is a register of missions, one row per mission
(Table~\ref{tab:register}). For each mission, this records the sending polity, the
receiving polity, the representative(s), the diplomatic rank, the dates of its
first and last recorded acts, the stated purpose where the entry gives one, and
the archival or published source. Representatives are matched to the Repertorium's
index of persons, so that one who served in more than one mission can be
identified across the register.

\begin{table}[!ht]
\centering\small
\caption{Variables of the mission register. Each row is one mission.}
\label{tab:register}
\begin{tabular}{@{}l p{0.66\textwidth}@{}}
\toprule
Variable & Description\\
\midrule
Mission id & Unique identifier for the mission\\
\addlinespace
Sending polity & The polity that sent the representative\\
\addlinespace
Receiving polity & The court or polity to which the representative was sent\\
\addlinespace
Representative(s) & The named representative or representatives, linked to the persons table\\
\addlinespace
Rank & The diplomatic rank, as printed and coded on five variables (Table~\ref{tab:ranks})\\
\addlinespace
First act & Date of the first recorded act, taken as the mission's start\\
\addlinespace
Last act & Date of the last recorded act, taken as the mission's end\\
\addlinespace
Purpose & The stated business of the mission, where the entry gives one\\
\addlinespace
Source & The archival or published citation\\
\addlinespace
Page & The page in Band~I on which the entry appears\\
\bottomrule
\end{tabular}
\end{table}

Rank is coded on five variables (Table~\ref{tab:ranks}): its ceremonial level
(ambassador, envoy or minister, resident, agent or commissary, secretary or
charg\'e, courier, or other); whether the post was resident or temporary; whether
the appointment was ordinary or extraordinary; whether the representative held
full powers (\emph{plena potestas}), that is, formal authority to negotiate and
sign an agreement on the sovereign's behalf, styled \emph{bevollm\"achtigt} or
\emph{pl\'enipotentiaire} in the source; and whether the rank belonged to the
papal series (nuncio, internuncio, cardinal, legate, and related Curial titles).
The original wording is kept alongside the coded values.

\begin{table}[!ht]
\centering\small
\caption{Coding of diplomatic rank. Each of the 158 distinct rank descriptions in
Band~I is coded on five variables.}
\label{tab:ranks}
\begin{tabular}{@{}l p{0.34\textwidth} p{0.40\textwidth}@{}}
\toprule
Variable & Description & Values\\
\midrule
Ceremonial level & The rank's position in the ceremonial hierarchy &
ambassador; envoy or minister; resident; agent or commissary; secretary or
charg\'e; courier; other\\
\addlinespace
Resident post & Whether the post was a standing residence or a temporary mission &
yes; no\\
\addlinespace
Extraordinary & Whether the appointment was styled extraordinary or ordinary &
extraordinary; ordinary; unspecified\\
\addlinespace
Full powers & Whether the representative held full powers (\emph{plena potestas})
to negotiate and sign agreements & yes; no\\
\addlinespace
Papal series & Whether the rank belongs to the Curia's own vocabulary (nuncio,
cardinal, legate, and related titles) & yes; no\\
\bottomrule
\end{tabular}
\end{table}

The register rests on an underlying event table, in which each row is a single
dated act, such as a presentation of credentials, an audience, a recall, or a
departure; the mission dates are taken from these acts. For comparison with the
annual series available after 1817 \citep{bayer2006}, I aggregate the register
into a dyad-year panel, in which each row is an ordered pair of polities in a given
year, with the number of active missions and the ranks present. Treaty signings
are kept separately.

\subsection{Validation}

I subject the automated extraction to three kinds of validity checks. First, I
apply internal-consistency checks derived from the structure of the Repertorium.
If extraction has been carried out correctly, a set of structural relations must
hold: each entry falls under exactly one sending chapter; the chapters together
cover the whole volume without overlapping; each name in the index appears on the
page the index lists for it; and a treaty that appears in more than one chapter
has the same date in each. These structural checks identified 492 entries that had
been assigned to the wrong sending chapter, which were subsequently corrected.

Second, I measure coverage against the Repertorium's index of persons, an
alphabetical index (compiled separately from the country chapters) that lists
every diplomat named in the volume together with the pages on which each appears.
Because it enumerates the same persons independently of the chapter parse, the
names it lists for a polity are the set the parse should have recovered, which
makes it a natural benchmark for recall. Exact surname matching recovers 6,156 of
6,656 indexed persons (92.5 percent). As an additional check, treaty signings
recorded in more than one chapter agree on the full date in 210 of 215 cases; the
five exceptions are discrepancies in the printed volume rather than extraction
errors.

Third, I assess accuracy by direct comparison with human coding. A stratified
human audit of a random sample of entries remains to be completed before data
release; it will report precision and recall separately for senders, receivers,
persons, ranks, dates, and citations.

\subsection{Limitations}

The Repertorium is likely the most encompassing source on diplomatic
representation in early modern Europe. Three limitations nonetheless deserve
mention. First, since each national chapter was compiled by a different
correspondent working the archives available in that country, coverage is uneven
across polities. The implication is that the Repertorium is more comprehensive
for some polities than for others. For instance, many correspondents worked on
the German Empire and the Habsburg lands, while Spain, Russia, and the Ottoman
Empire had few or none \citep{bittnergross1936, jablonowski1938}.

Second, the coverage of personnel is selective. The editors report that they
focused on the permanently accredited heads of mission and their charg\'es
d'affaires \citep[p.~VIII]{bittnergross1936}, later adding special missions and
treaty negotiators without a claim to completeness \citep{jablonowski1938}. The
8,852 person records here therefore under-represent the subordinate staff who
served in European legations. The register nonetheless covers most fully the
representatives most relevant to research, the heads of mission who directed each
legation and who are the usual unit of analysis in the study of diplomatic
representation.

Third, the dates are censored at the boundaries of the period. Missions already
active in 1648, or still active in 1715, are recorded only for the part of their
span that falls within it, though dates outside 1648--1715 are retained where the
volume reports them. The two calendars then in use, the Julian and the Gregorian,
are less of a concern: the Repertorium follows a stated convention, giving dual
dates as Julian first and Gregorian second and treating any unmarked date as New
Style (Gregorian), and it flags the 127 genuinely doubtful cases in the text.

\section{Descriptive patterns, 1648--1715}

\begin{figure}[!ht]
\centering
\includegraphics[width=\textwidth]{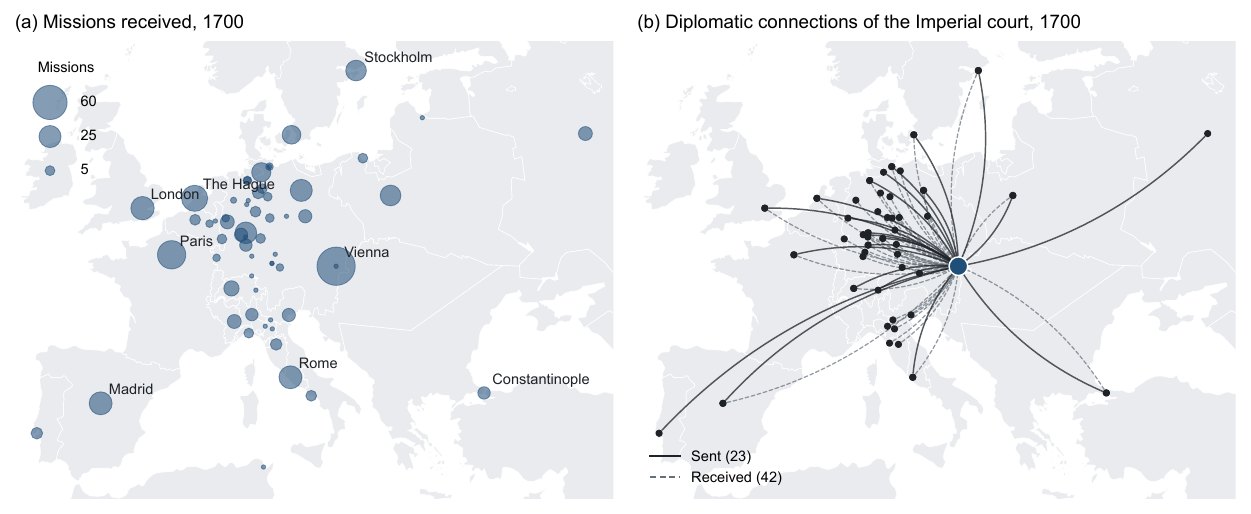}
\caption{Diplomatic representation in 1700. Panel (a) plots receiving polities,
with marker size proportional to the number of missions present. Panel (b)
shows missions sent and received by the Imperial court in Vienna. The borders
are an approximate historical base map for c.\ 1700, from the
\emph{historical-basemaps} project (\url{https://github.com/aourednik/historical-basemaps};
after Euratlas and N\"ussli, CC BY-SA 4.0).}
\label{fig:map}
\end{figure}

Figure~\ref{fig:map} gives a snapshot of the data in a single year, 1700. In
panel (a), each point marks a polity that received diplomatic representation, and
its size is proportional to the number of missions present there that year. A few
polities, including the period's great powers and important courts, attracted many
missions, while a much larger number of smaller polities, including many in the
German states, received a smaller number. Panel (b) illustrates the networked
nature of the data, showing the in- and outgoing diplomatic connections of the
Emperor's court in 1700. The incoming connections (42) outnumber the outgoing
(23), and the incoming side is dominated by the smaller German principalities and
the Italian states, while the Emperor's own missions focused on the major powers.

Table~\ref{tab:top} reports the count of sent and received missions. The Emperor
ranks first in both categories. France and Brandenburg-Prussia occupy the next two
positions, although their order differs between missions sent and received.
Outside Europe, the data record missions to the Ottoman Empire (296), Persia (29),
and the Barbary states of Tunis (18) and Algiers (15).

\begin{table}[!ht]
\centering\small
\caption{Senders and receivers most frequently recorded, 1648--1715.}
\label{tab:top}
\setlength{\tabcolsep}{4.5pt}
\begin{tabular}{lrlr}
\toprule
Sender & Missions & Receiver & Missions\\
\midrule
Holy Roman Empire, Emperor & 1,188 & Holy Roman Empire, Emperor & 1,355\\
France & 982 & Brandenburg-Prussia & 880\\
Brandenburg-Prussia & 882 & France & 750\\
England & 609 & Imperial Diet at Regensburg & 728\\
Poland & 488 & England & 653\\
Mainz & 459 & Dutch Republic & 625\\
Sweden & 401 & Poland & 422\\
Denmark & 381 & Sweden & 415\\
\bottomrule
\end{tabular}
\end{table}

\begin{figure}[!ht]
\centering
\includegraphics[width=.8\textwidth]{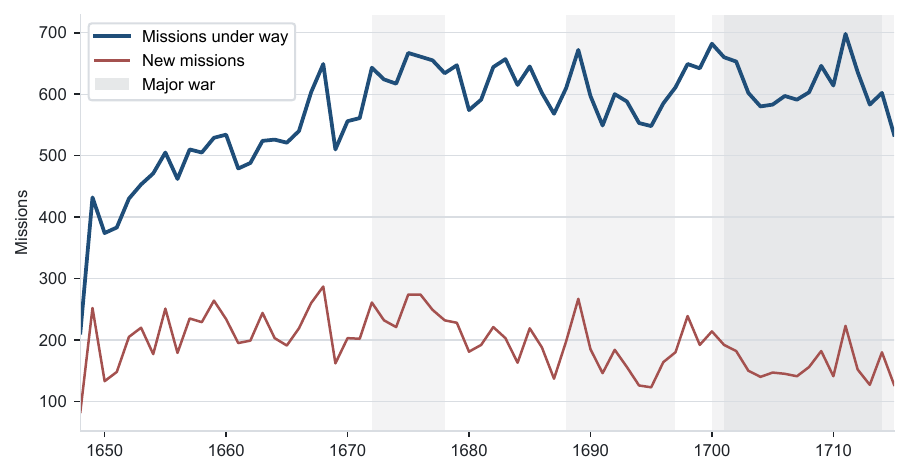}
\caption{Active and newly recorded missions by year, 1648--1715.}
\label{fig:system}
\end{figure}

Turning to over-time patterns, Figure~\ref{fig:system} shows the number of active
missions each year. Activity rises from about 200 missions in the late 1640s to an
average of 537 in the 1660s. It then remains broadly stable, generally between 550
and 700 missions, and reaches a peak of 698 in 1711. The shaded intervals identify
the major wars of the period, including the Franco-Dutch War, the Nine Years' War,
the War of the Spanish Succession, and the Great Northern War. None coincides with
a clear and sustained increase or decline in representation.

\begin{figure}[!ht]
\centering
\includegraphics[width=\textwidth]{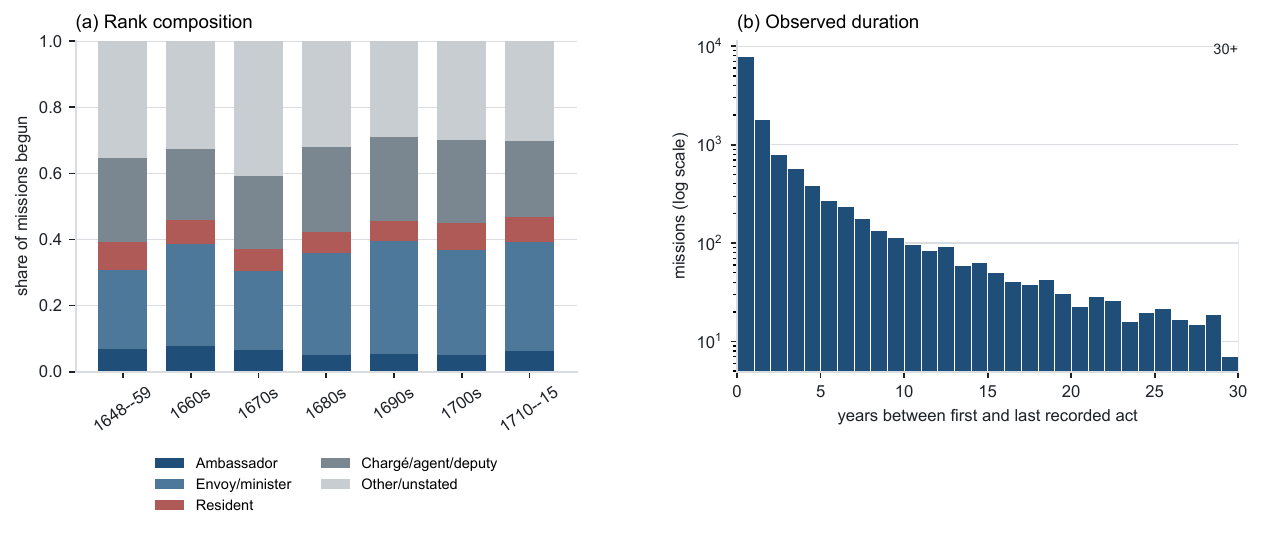}
\caption{Rank and duration. Panel (a) gives the rank composition of missions by
decade of first recorded act. Panel (b) gives the observed number of years
between the first and last act.}
\label{fig:missions}
\end{figure}

Figure~\ref{fig:missions}a reports the ranks of missions first observed in each
decade. Envoys and ministers grew to be the largest category over the period,
while ambassadors remained a small minority, even in the later decades. A
substantial share of missions, especially in the early decades, carry no stated
rank.

Figure~\ref{fig:missions}b reports the observed duration of missions, measured
from the first to the last recorded act. Most are short, with 79 percent lasting
no more than two years. At the other end of the distribution, a small number of
missions continue for a decade or longer. These long spans include residents and
other representatives attached to continuing posts.

The data also support analysis of individual representatives. They contain 8,852
person records, of which 2,274 are linked to more than one mission and 446 to more
than one sending polity, suggesting that a sizeable set of individuals took on
diplomatic representation tasks for more than one ruler during their career. Six of
the representatives are women.\footnote{Five were noblewomen sent by France and one
by the Palatinate, in extraordinary ceremonial ranks (\emph{ambassadrice
extraordinaire}, \emph{envoy\'ee}, or plenipotentiary). Where their purpose is
recorded, it was to escort a bride or princess to a foreign court: the princesse de
Lillebonne, for example, accompanied Anne-Marie d'Orl\'eans to Savoy for her
marriage in 1684.}

\section{An illustrative application: ruler succession and diplomatic
continuity}

To illustrate the analytical value of the mission dates, I examine whether ruler
succession impacts the rotation of diplomatic missions. This makes it possible to
distinguish whether the diplomatic apparatus of this period was patrimonial, that
is, tied to the ruler, or an impersonal organization less sensitive to leader
deaths or exits \citep{ertman1997, nww2009}. Previous studies have used leader
deaths to estimate leader effects and political turnover to study bureaucratic
persistence \citep{jonesolken2005, iyermani2012, akhtarimoreiratrucco2022}.

The analysis covers 63 ruler transitions between 1651 and 1712, coded from
secondary sources and verified against data on rulers collected by
\citet{kokkonen2022}. For each transition, I identify the polity's missions active
in that year. The termination rate is the share of these missions whose last
recorded act occurred in the same or following year. I compare this rate with the
corresponding rate for the same polity in ordinary years, defined as years outside
the three-year window surrounding any of its ruler transitions. The difference can
be called the succession excess.

\begin{figure}[!ht]
\centering
\includegraphics[width=.82\textwidth]{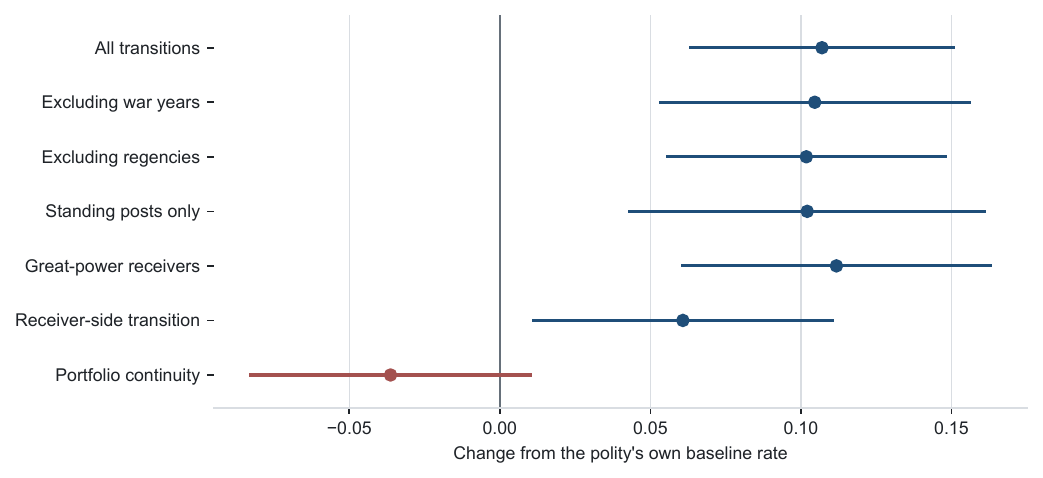}
\caption{Succession excess across specifications. Each estimate is the change in
a polity's mission-termination rate at a ruler transition, relative to its own
ordinary-year rate, with 95 percent confidence intervals. The bottom row reports
the change in portfolio continuity, the share of receiving polities retained
across a transition.}
\label{fig:succession}
\end{figure}

Figure~\ref{fig:succession} shows that diplomatic missions ended more frequently
around ruler transitions. The termination rate was 10.7 percentage points higher
in transition years than in ordinary years (SE $= 2.2$).\footnote{The reported
standard errors treat the 63 transitions as independent. Clustering them by the 25
polities does not substantively change the result and, if anything, reduces the
uncertainty (clustered standard error 1.9 percentage points, $t=5.7$).} In a
permutation test that assigns each transition to a randomly selected ordinary year
for the same polity, none of the simulated estimates was as large as the observed
estimate ($p<.003$). Excluding transitions that fell in the onset or termination
year of a major war leaves the estimate unchanged, at 10.5 percentage points (SE
$= 2.6$). The estimates are also similar when regencies (successions in which a
regent governed for a minor or incapacitated ruler) are excluded and when the
analysis is restricted to standing posts or great-power receivers.

\begin{figure}[!ht]
\centering
\includegraphics[width=.72\textwidth]{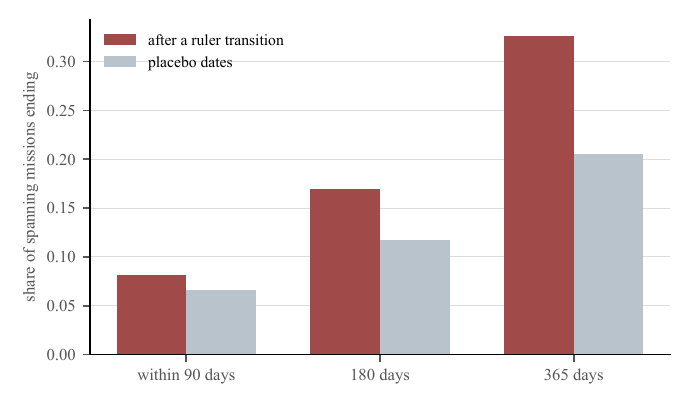}
\caption{Additional test at day resolution. The figure reports the share of
missions active on the transition date that ended within 90, 180, and 365 days,
compared with placebo dates for the same polity, over the 40 day-dated
transitions.}
\label{fig:day}
\end{figure}

The day-level data show the same temporal pattern. Among the 40 transitions dated
to the day, 32.6 percent of the missions active on the transition date ended within
the following year, against 20.6 percent at placebo dates for the same
polity\footnote{For each transition I draw 20 random placebo dates for the same
polity, each at least two years from any transition. The shares count missions
active on the date that lasted at least 60 days, and are computed only for
transitions with at least three such missions.} (Figure~\ref{fig:day}).

There is no clear evidence that the association weakened during the period. The
estimated succession excess was 12.2 percentage points before 1683 and 9.1 points
thereafter. This suggests that diplomatic appointments remained tied to the ruler
throughout the period, rather than becoming more impersonal. Given the relatively
small number of ruler transitions per polity, there is not sufficient power to
detect variation across polities.

While ruler succession affected who held a diplomatic appointment, the findings
indicate no detectable effect on the portfolio of diplomatic partners, that is, the
polities with which a polity engaged diplomatically. I measure portfolio continuity
as the share of receiving polities served during the three years before a
transition that were also served during the three years after it. Portfolio
continuity was 3.6 percentage points lower at successions than in ordinary years
(SE $= 2.4$), a difference that is not statistically distinguishable from zero.
Succession thus affected individual mission records more strongly than the set of
diplomatic relationships maintained by a polity.

\section{Conclusion}

This paper has introduced a dataset of diplomatic representation in Europe for
1648--1715, drawn from Band~I of the \emph{Repertorium der diplomatischen
Vertreter aller L\"ander seit dem Westf\"alischen Frieden}, a reference work
cataloguing the diplomatic representatives exchanged among European polities
since the Peace of Westphalia. The data provide a quantitative view of a large
and heterogeneous system, with 141 sending polities represented at 201 receiving
polities across 1,796 directed pairs. They begin over a century and a half before
the earliest comparable existing quantitative series \citep{bayer2006,
teorell2023} and, in many cases, record diplomatic missions at a level of detail
and temporal granularity not captured in datasets covering more recent eras.

The descriptive patterns indicate that diplomatic activity was already extensive
by the late seventeenth century, concentrated on a few major courts but reaching
a large number of smaller polities. Residents made up about a fifth of active
representation, and their share did not rise over the period, qualifying accounts
of the emergence of the resident embassy.

An illustrative application showed that diplomatic missions ended more frequently
when the sending polity changed rulers, suggesting that appointments were tied to
the person of the ruler. The set of partners a polity served, however, changed
little across successions. Ruler transitions therefore affected individual
missions more strongly than diplomatic portfolios. This pattern is consistent with
ruler-specific commissions operating within durable relationships between polities.

Several steps remain before the data are complete and can be released. A human
transcription audit is a first priority, to estimate precision and recall for the
automated extraction, alongside extending the data forward to 1815.

\bibliographystyle{apalike}
\bibliography{refs}

\end{document}